\documentclass[12pt,amsmath,amssymb,superscriptaddress]{revtex4}

\usepackage{graphicx}
\usepackage{dcolumn}
\usepackage{bm}
\usepackage{color}
\usepackage{textcomp}

\begin{document}


\title{Free-Electron Laser-Powered Electron Paramagnetic Resonance Spectroscopy} 

\author{S. Takahashi}
\email{susumuta@usc.edu}
\affiliation{Department of Chemistry, University of Southern California, Los Angeles CA 90089, USA}%
\affiliation{Institute for Terahertz Science and Technology, University of California, Santa Barbara CA 93106, USA}%
\author{L.-C. Brunel}
\affiliation{Institute for Terahertz Science and Technology, University of California, Santa Barbara CA 93106, USA}%
\author{D. T. Edwards}
\affiliation{Department of Physics, University of California, Santa Barbara CA 93106, USA}%
\author{J. van Tol}
\affiliation{National High Magnetic Field Laboratory, Florida State University, Tallahassee FL 32310, USA}%
\author{G. Ramian}
\affiliation{Institute for Terahertz Science and Technology, University of California, Santa Barbara CA 93106, USA}%
\author{S. Han}
\affiliation{Institute for Terahertz Science and Technology, University of California, Santa Barbara CA 93106, USA}%
\affiliation{Department of Chemistry and Biochemistry, University of California, Santa Barbara CA 93106, USA}%
\author{M. S. Sherwin}
\affiliation{Institute for Terahertz Science and Technology, University of California, Santa Barbara CA 93106, USA}%
\affiliation{Department of Physics, University of California, Santa Barbara CA 93106, USA}%

\date{\today}

\maketitle

{\bf
Electron paramagnetic resonance (EPR) spectroscopy interrogates unpaired electron spins in solids and liquids to reveal local structure and dynamics; for example, EPR has elucidated parts of the structure of protein complexes that have resisted all other techniques in structural biology~\cite{hubbell96, hubbell00, pannier00, saxena96, borbat99}. EPR can also probe the interplay of light and electricity in organic solar cells~\cite{Smilowitz93, Dyakonov99, Ogiwara05} and light-emitting diodes~\cite{mccamey08}, and the origin of decoherence in condensed matter, which is of fundamental importance to the development of quantum information processors~\cite{ takahashi08, Bruber96, lyon06, hanson07, takahashi11}. Like nuclear magnetic resonance (NMR), EPR spectroscopy becomes more powerful at high magnetic fields and frequencies, and with excitation by coherent pulses rather than continuous waves. However, the difficulty of generating sequences of powerful pulses at frequencies above 100 GHz has, until now, confined high-power pulsed EPR to magnetic fields of 3.5 T and below. Here we demonstrate that $\sim$1 kW pulses from a free-electron laser (FEL) can power a pulsed EPR spectrometer at 240 GHz (8.5 T), providing transformative enhancements over the alternative, a state-of-the-art $\sim$30 mW solid state source. Using the UC Santa Barbara FEL as a source, our 240 GHz spectrometer can rotate spin-1/2 electrons through $\pi$/2 in only 6 ns (vs. 300 ns with the solid state source).  Fourier transform EPR on nitrogen impurities in diamond demonstrates excitation and detection of EPR lines separated by $\sim$200 MHz.  Decoherence times for spin-1/2 systems as short as 63 ns are measured, enabling measurement of the decoherence time in a frozen solution of nitroxide free-radicals at temperatures as high as 190 K. Both FELs and the quasi-optical technology developed for the spectrometer are scalable to frequencies well in excess of 1 THz, opening the possibility of high-power pulsed EPR spectroscopy up to the highest static magnetic fields on earth.
}


The spectral resolution, spin polarization, sensitivity, and time resolution of pulsed EPR all increase with increasing static magnetic field and the associated Larmor precession frequency (Fig. 1a)~\cite{freed00}. An additional premium to high magnetic fields has recently been demonstrated at 8.5 T ($f_{Larmor}$=240 GHz): because of the 11.5 K Zeeman energy, the spin polarization varies from $>$99$\%$ at 2 K to $<$2$\%$ at 300 K, so that the spin decoherence time $T_2$ increases dramatically with decreasing temperature in a large class of spin systems~\cite{takahashi08, takahashi09fe8}.
In addition to the advantages of high fields and frequencies, pulsed EPR benefits from high microwave power. In a fixed geometry, the time required to rotate an ensemble of spins by a given angle, say $\pi/2$, is proportional to the strength of the resonant microwave magnetic field $B_1$, and hence to the square root of the power. For many systems of interest --- for example, spin-labeled proteins above the 200 K protein glass transition, where they explore their biologically-relevant conformational space, or spins in semiconductors (and hence electronic devices) above 20 K --- spin relaxation times are much shorter than $\sim$1 $\mu$s (for spin 1/2 systems)~\cite{earle05, vantol09}, and cannot be measured using state-of-the-art solid state sources at 240 GHz which typically have powers of $\sim$30 mW. Thus, it is imperative to have much higher microwave power allowing $\pi/2$ spin rotations in only a few nanoseconds.

 Electromagnetic sources are the bottleneck in the development of high-power pulsed EPR at high magnetic fields. Most high-power pulsed EPR spectrometers in the world operate at 0.34 T ($f_{Larmor}$=9.5 GHz) or 1.2 T ($f_{Larmor}$=34 GHz) and are powered by vacuum electronic devices called traveling wave tube amplifiers~\cite{brukere580}.  The highest magnetic fields at which spectrometers with few-ns $\pi/2$ pulses have been constructed is 3.5 T ($f_{Larmor}$=95 GHz). The 95 GHz instruments, pioneered at Cornell University and, more recently, beautifully implemented at the University of St. Andrews, UK, are powered by more exotic vacuum electronic devices called extended-interaction klystron amplifiers with peak output powers of $\sim$1 kW~\cite{hofbauer04, cruickshank09}. Amplifiers capable of kW output powers at frequencies above 200 GHz have been envisioned~\cite{cpi, hall97, fukuiTHz} but are beyond current technology.

 In this Report, we present results from a pulsed EPR spectrometer powered by a free-electron laser (FEL). The spectrometer operates at 8.5 T (240 GHz), and can be powered either by a 30 mW, 240 GHz solid-state source or by UCSB's millimeter-wave FEL (mm-FEL)~\cite{ramian92}, with maximum power of 1 kW.

\begin{figure}
\includegraphics[width=150 mm]{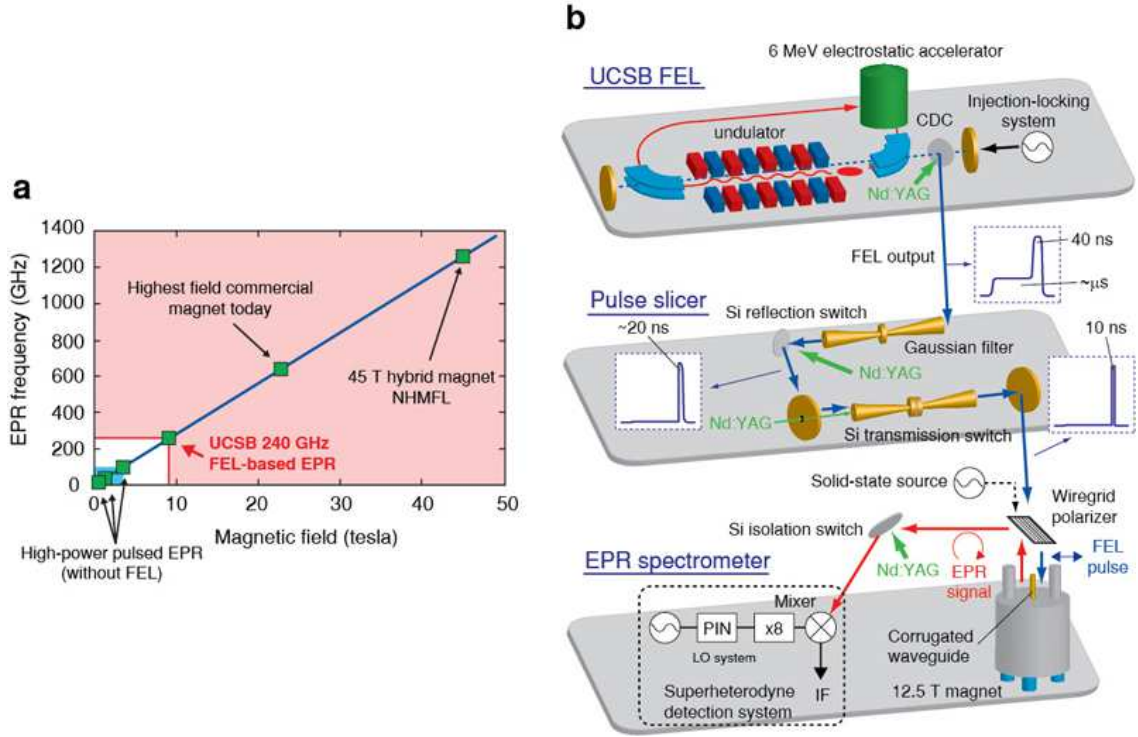}
\caption{\label{Fig1}
(a) Larmor precession frequencies and their corresponding magnetic fields for the electron. The highest d.c. magnet is available at the national high magnetic field laboratory (NHMFL). (b) Schematic overview of the FEL powered pulsed EPR system.
The UCSB millimeter-wave FEL emits a several microsecond-long pulse. For control of frequency and output power, an injection-locking system and a cavity-dump coupler (CDC) have been implemented. The Nd:YAG laser-controlled pulse-slicer functions to "slice" one or two pulses, each with variable duration as short as 1 ns, from the FEL radiation. The pulsed EPR spectrometer operates quasi-optically. For the cw and low-power EPR mode, the system is operated with a solid-state source at 240 GHz (Virginia Diode Inc.). The linearly-polarized output of the pulse-slicer is propagated into the center of the 12.5 T superconducting magnet in the EPR spectrometer where the nanosecond FEL pulses couple to the sample. The EPR signals are detected by a superheterodyne detection system after turning ON the Si isolation switch which protects the mixer during the pulse.
}
\end{figure}


An overview of the UCSB FEL-powered pulsed EPR system is shown in Fig.~1b (see the supporting online materials for details of the setup and the performance).
The UCSB mm-FEL~\cite{ramian92} emits few-microsecond long pulses with several hundred watts to several kilowatts peak power. In 240 GHz pulsed EPR operation, the FEL output is injection-locked by a stable 240 GHz solid-state source to generate few-microsecond long  pulses with a Fourier-transform limited linewidth $<$ 1 MHz~\cite{takahashi07}. By cavity-dumping the FEL, the several-hundred-Watt pulses can be stepped up to 1 kW for the last 40 ns, with $\sim$10 ns rise and fall times (Fig. 1b)~\cite{takahashi09}.

In the FEL-powered spectrometer, light-activated shutters "slice" one or two pulses directly from the $>$100 W beam, each with a variable duration as short as 1 ns. This is a very different mode of operation than is used in other high-power pulsed EPR spectrometers, where pulse sequences are generated at low power and then amplified. The shutters are high-purity Si wafers with thicknesses carefully polished down to 1/2 wavelength~\cite{hegmann96, doty04}. In our "pulse slicer" (see Fig. 1b and the supporting online materials), "reflection switches" are Si wafers at Brewster's angle which reflect less than 1 part per million in the quiescent state.  When excited with 532 nm (green) light of a Nd:YAG laser with a fluence $>$ 1 mJ/cm$^2$, the reflectance rises to $>$60$\%$  with a sub-ns rise time. The reflectance remains high for at least $\sim$1 $\mu$s. To make a pulse of variable duration, the 240 GHz beam is directed from a reflection switch into a "transmission switch" consisting of back-to-back corrugated horns with a Si wafer between them, illustrated in Fig. 1b. For a single assembly, the transmittance drops from near unity in the quiescent state to less than 10$^{-4}$ when the Si is illuminated with a pulse of green light. The duration of the pulse exiting the transmission switch is determined by the optical or electronic delay between the green pulses which activate the reflection and transmission switches. Transmittance switches are cascaded to achieve transmittances well below 10$^{-6}$. A non-trivial arrangement of reflection and transmittance switches enables the generation of pairs of pulses separated by a time variable from a few ns up to the several $\mu$s duration of the FEL pulse. This Si switch technology is broadband, and has been demonstrated up to frequencies of at least 30 THz~\cite{rolland86}. A pulse slicer similar to the one developed here could be used for EPR with other quasi-cw sources like Gyrotrons~\cite{mitsudo10}.

The output of the pulse slicer, or of a 30 mW solid state source, enters a quasi-optical spectrometer very similar in design to systems that have been described previously~\cite{smith98, freed00, vantol05}. After passing through a wiregrid polarizer, 240 GHz pulses travel down an 18 mm corrugated waveguide to a taper that reduces the beam size to either 5 or 2 mm. Solid samples are mounted directly on top of a silver coated mirror and frozen liquid samples contained in a small teflon bucket on the mirror. The samples are then placed at the exit of the taper in a 12.5 T superconducting magnet. No resonator is used. The 240 GHz pulse tips the spins that precess at 240 GHz, to the plane perpendicular to a static magnetic field, resulting in the emission of a sub-microWatt circularly polarized signal component that travels back up the corrugated waveguide following the $>$100 W level excitation radiation. The superheterodyne receiver is isolated from the excitation pulse and subsequent reflections by the induction mode detection~\cite{smith98} that only guides the EPR signal component to the detector using the wiregrid polarizer ($\sim$0.3 $\%$ reflection for FEL pulse, 50$\%$ for EPR signal) and the isolation switch (1 ppm reflection before activation by a laser pulse).


\begin{figure}
\includegraphics[width=150 mm]{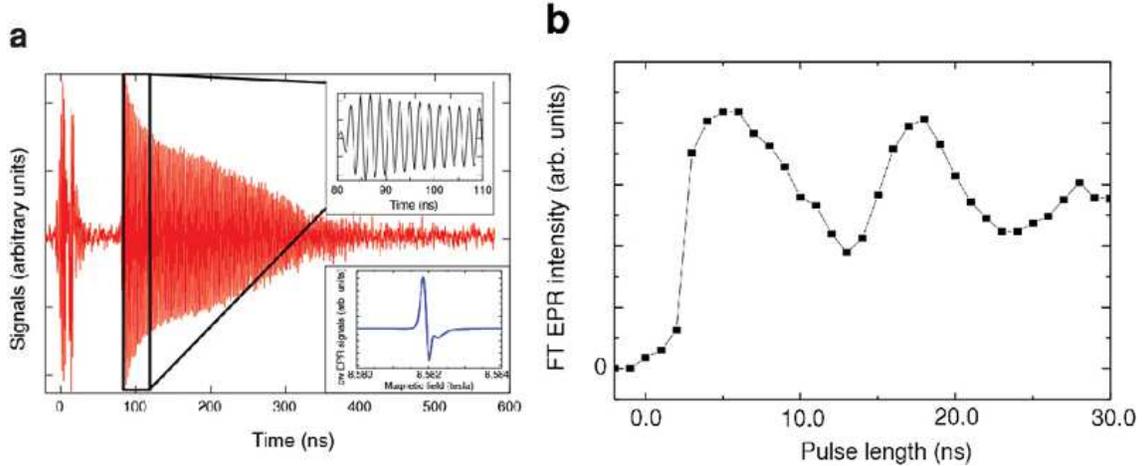}
\caption{\label{Fig2} FEL-powered EPR measurements in BDPA. (a) Free-induction decay signals of BDPA crystals. The intermediate frequency (IF = 500 MHz) signals in the heterodyne detection system were recorded by a fast digitizer to observe the FID. The spectrum was taken at room temperature with a single pulse and scan. The receiver is turned ON at 80 ns. The inset shows a 0.3 mT wide cw EPR spectrum of BDPA. (b) Rabi oscillations of BDPA acquired by FT-EPR at room temperature. The intensity of the FT-EPR signals was measured as a function of the length of a single pulse with a single scan. The damping of the observed oscillations is due to inhomogeneous distribution of $B_1$  intensity over the sample.
}
\end{figure}

In order to characterize the FEL-powered spectrometer, we used a 1:1 crystalline complex of $\alpha$, $\gamma$-Bisdiphenylene-$\beta$-phenylallyl (BDPA) and benzene. This stable radical is widely used in EPR and related techniques. The sample dimension is much smaller than the 1.25 mm wavelength of the transmitted radiation, and the exchange-narrowed 0.3 mT wide EPR signal was measured by cw EPR using a solid state source.  The magnetic field was set to the center of the EPR peak.  A 10 ns pulse sliced from the FEL was applied and, after 80 ns, the isolation switch was activated.  What is observed is the decaying oscillation of the "free-induction decay" (FID) of BDPA, mixed down to an intermediate frequency (IF) measurable by a transient digitizer (Fig. 2a), {\it i.e.} IF=500 MHz. The intensity of the FID is then recorded by taking a fast-Fourier transform (FFT) of the FID and determining the signal intensity. In Fig. 2b, the signal intensity is plotted as a function of pulse duration varying from 0 to 30 ns in 1 ns increments. The signal undergoes "Rabi oscillations"~\cite{rabi32}, rising to its first maximum near 5 ns pulse duration.  At the first maximum, most spins have rotated by $\pi$/2. The frequency of oscillation is $\sim$40 MHz, corresponding to a $\pi$/2 pulse of $\sim$6 ns and a transverse magnetic field $B_1$ of $\sim$3.0 mT at the sample.  The minima of the Rabi oscillations at 12 and 25 ns are not zero, indicating a significant inhomogeneity in $B_1$. The origin of this inhomogeneity is under investigation. The 6 ns $\pi$/2 pulse is 50 times shorter than can be achieved in this and similar spectrometers for a spin-1/2 system with a lower power solid-state source.

\begin{figure}
\includegraphics[width=150 mm]{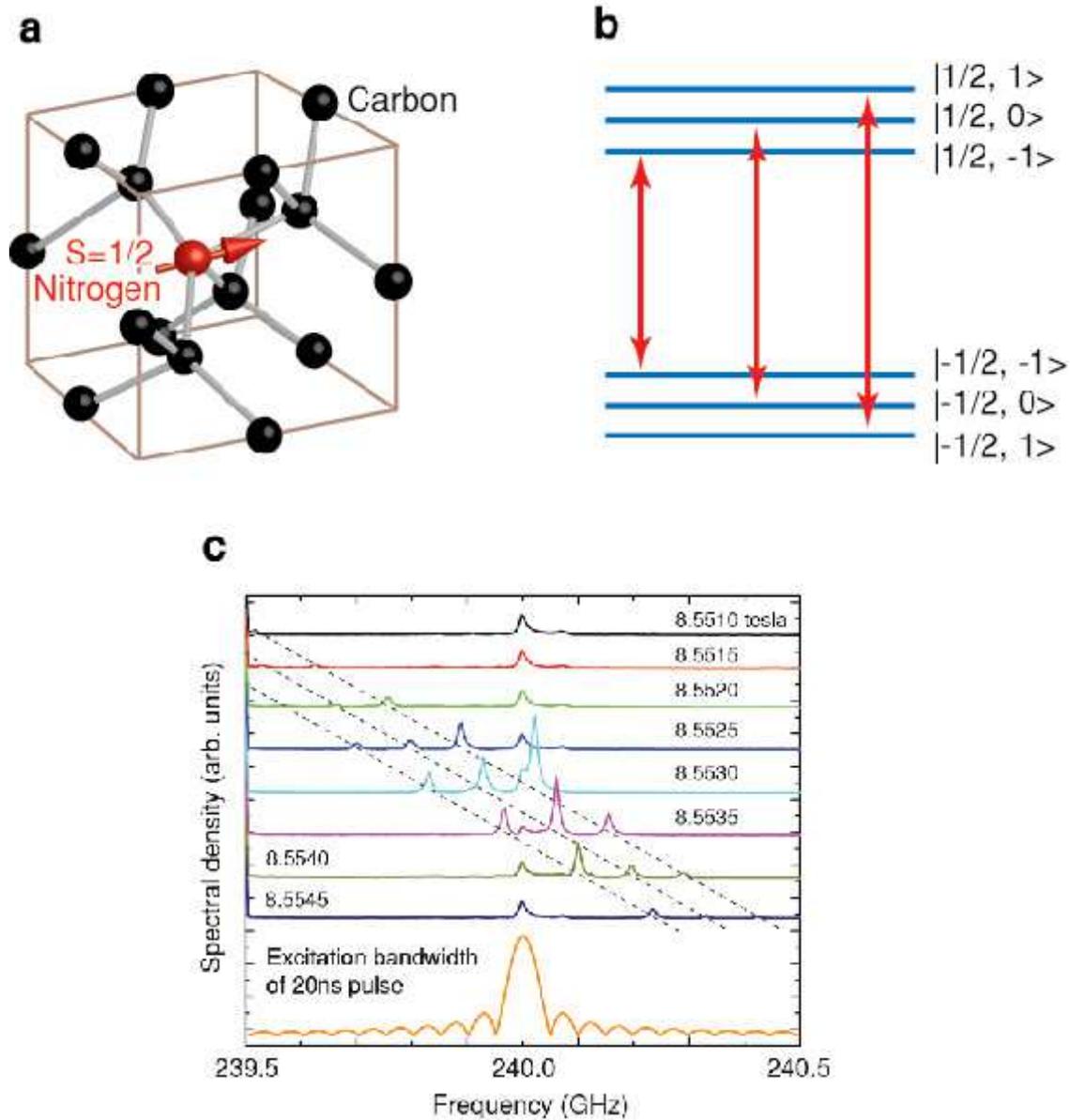}
\caption{\label{Fig3} (a) A single nitrogen impurity in a diamond lattice. The nitrogen has an unpaired electron spin ($S=1/2$). (b) Due to hyperfine couplings to $^{14}$N nuclear spin, the energy states of the nitrogen electron spin are split into 6 states in a high magnetic field and three pronounced EPR lines are observed. The spin states are represented by $|m_S, m_I>$ where $m_S$ and $m_I$ are magnetic quantum numbers of electron and nuclear spins respectively. (c) Fourier-transform EPR of single nitrogen impurities in diamond taken at different magnetic fields. The spectra were taken at room temperature with 60 averages. The graph showed three EPR peaks because of hyperfine couplings between N electron and $^{14}$N nuclear spins. The peaks at 240 GHz, that are independent of the intensity of the magnetic fields, are due to leakage of the FEL radiation. The calculated excitation bandwidth of a square 20 ns pulse is shown at the bottom of the figure.
}
\end{figure}

The short excitation pulses enable Fourier Transform EPR of single nitrogen (N) impurity spins in diamond. As shown in Fig.~3a, a N impurity exists in a diamond lattice as a substitutional impurity and has one unpaired electron with spin S=1/2.
Because of the hyperfine coupling between the N electron spin and the $^{14}$N nuclear spin (I=1),
the energy levels of the N electron spin are split into six levels, as described in Fig.~3b. Therefore three EPR peaks are observed from the N electron spins in a single crystal of diamond with the application of magnetic fields along the (100) direction.
Fig.~3c shows FT-EPR measurements performed on N spins at room temperature at different magnetic fields. We applied a single FEL pulse with 20 ns duration to perform FT-EPR. As shown in Fig.~3c, the observed three N EPR peaks were shifted by varying the magnetic fields.
The separation between neighboring peaks is 94 MHz, the expected hyperfine splitting for $^{14}$N impurities in diamond. Fig.3c shows that the FT-EPR spectrum acquired by the FEL-powered pulsed EPR spectrometer is clearly observable over more than 200 MHz bandwidth. This wide excitation bandwidth is necessary to manipulate all three N spin EPR transitions simultaneously --- more than 50 times broader than the 3 MHz bandwidth when we employ a lower power solid-state source.

\begin{figure}
\includegraphics[width=160 mm]{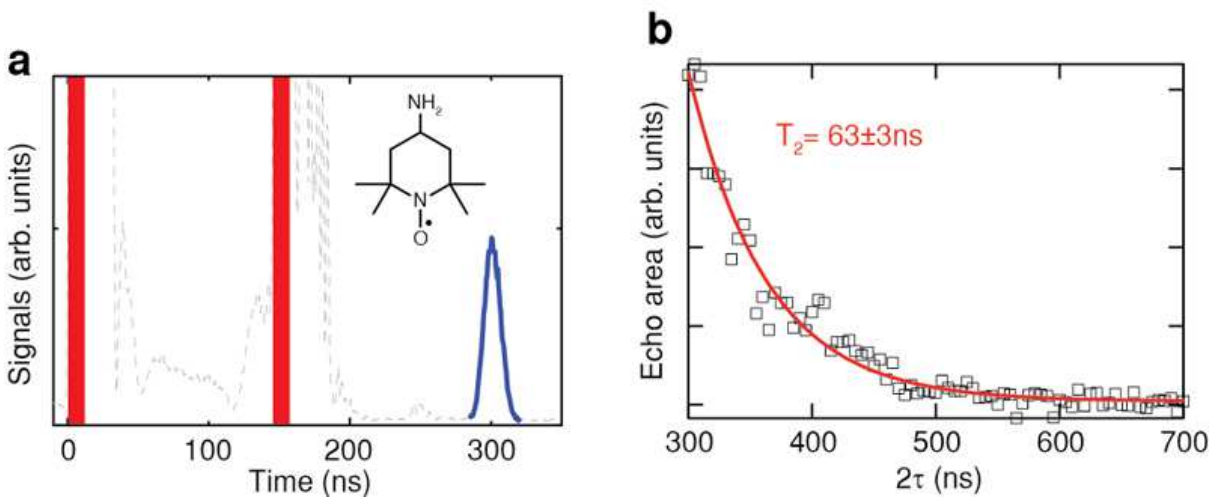}
\caption{\label{fig:Echo} (a) Hahn echo sequence of a nitroxide spin label showing two 10 ns sliced pulses (represented by red bars) followed by the echo signal (represented by a blue line) taken at 190 K with 7 averages. The receiver is tuned ON at 250 ns. Signal prior to that (dashed line) is from incomplete isolation. The inset shows a TEMPO molecule. (b) The echo area (the area under the blue line in Fig.~4a) is plotted as a function of twice the inter-pulse spacing, showing an echo decay over several hundred nanoseconds. The red line shows a fit to an exponential decay with $T_2 = 63$ ns.}
\end{figure}

Although single pulse experiments probe a wealth of information, magnetic resonance regularly utilizes multiple coherent pulses to perform more advanced experiments. The simplest example is the powerful Hahn echo sequence, which uses a second pulse to rotate the spins by 180\textdegree ~and refocuses the spins, eliminating the effects of static inhomogeneities in the local fields. Thus, echo experiments can allow for the detection of EPR signals in samples where inhomogeneous broadening causes a rapidly decaying FID, and can also be used to determine the spin decoherence time $T_2$ of a system. Echo signals were acquired on 4-amino-2,2,6,6-tetramethylpiperidine-1-oxyl (TEMPO) stable radicals dissolved in a deuterated glass composed of a frozen mixture of deuterated glycerol and D$_2$O (the volume ratio of 60:40). TEMPO is often used as a "spin label" in EPR studies of structure and dynamics of biological molecules. The sample was roughly 10 uL of frozen TEMPO solution of 50 mM radical concentration in a teflon bucket. Fig.~4a shows the two pulses used to generate the echo, and the echo signal itself for a single inter-pulse spacing of 150 ns at 190 K. A traditional experiment to determine the spin decoherence time was carried out by recording the decay of the echo area with increasing inter-pulse spacing (Fig.~4b). The pulses used are only $\sim$ 10 ns long. These short pulses make dramatically shorter relaxation times accessible, increase the bandwidth, and improve the signal-to-noise ratio by exciting more spins than the longer and weaker pulses available from a solid-state source. The spin echo shown in Fig.~4b measured at 190 K shows a single exponential decay with a time constant of 63$\pm$3 ns, an order of magnitude faster than what can be resolved for a spin-1/2 electron with our EPR spectrometer using a solid-state source. This enables $T_2$ measurements of nitroxide samples to be carried out at 190 K, nearly 100 K higher than is possible with a low-power solid-state source, and very close to the "glass transition" above which proteins become flexible and are functional. Currently, the inter-pulse spacing is limited by the spectrometer deadtime, where the residual ringdown in the spectrometer obscures the echo signal at short inter-pulse spacings.
To enable the detection of echo signals with inter-pulse spacings comparable to the short $\pi$/2 and $\pi$ pulses we have achieved, we are working to reduce the dead time by minimizing spurious reflections in corrugated waveguides and quasi-optics, and improving the isolation in our induction-mode detection following methods outlined in Ref. 21.


In summary, we have demonstrated FEL-powered pulsed EPR measurements.
The demonstration showed that the FEL-powered high-frequency EPR can manipulate S=1/2 spins with 6 ns $\pi/2$ pulses,
excite spin systems with $\sim$200 MHz bandwidth, and perform multipulse measurements such as the Hahn echo sequence. Methods to control the relative phase of a sequence of FEL pulses at 240 GHz, allowing for signal averaging and pulse/phase cycling, as well as methods for generating more sophisticated pulse sequences are under way. FELs only become more powerful and easier to use with increasing frequency, and the pulse slicing technology demonstrated here is scalable to much higher frequencies. Our spectrometer thus shows that pulse-sliced FELs can break through the bottleneck that has thus far limited high-power pulsed EPR to frequencies below 100 GHz. There are no obvious technological barriers to a FEL-based pulsed EPR spectrometer operating above 1 THz, and thus taking advantage of the highest static magnetic fields on earth to realize the full potential of EPR spectroscopy.


\end{document}


{\center{\large{\bf SUPPORTING ONLINE MATERIAL}}}

\vspace{3mm}

\title{Free-Electron Laser-Powered Electron Paramagnetic Resonance Spectroscopy}

\author{ S. Takahashi, L.-C. Brunel, D. T. Edwards, J. van Tol, G. Ramian, S. Han and M. S. Sherwin }

\makeatletter
\renewcommand{\thefigure}{S\@arabic\c@figure}


\maketitle

The high-powered pulsed EPR spectrometer at UCSB is constructed with a combination of a free-electron laser, a pulse slicer, and a quasioptical EPR bridge.
Here we provide details of the EPR spectrometer.

\vspace{3mm}

{\bf (i) UCSB FEL:}

\begin{figure}
\includegraphics[width=120 mm]{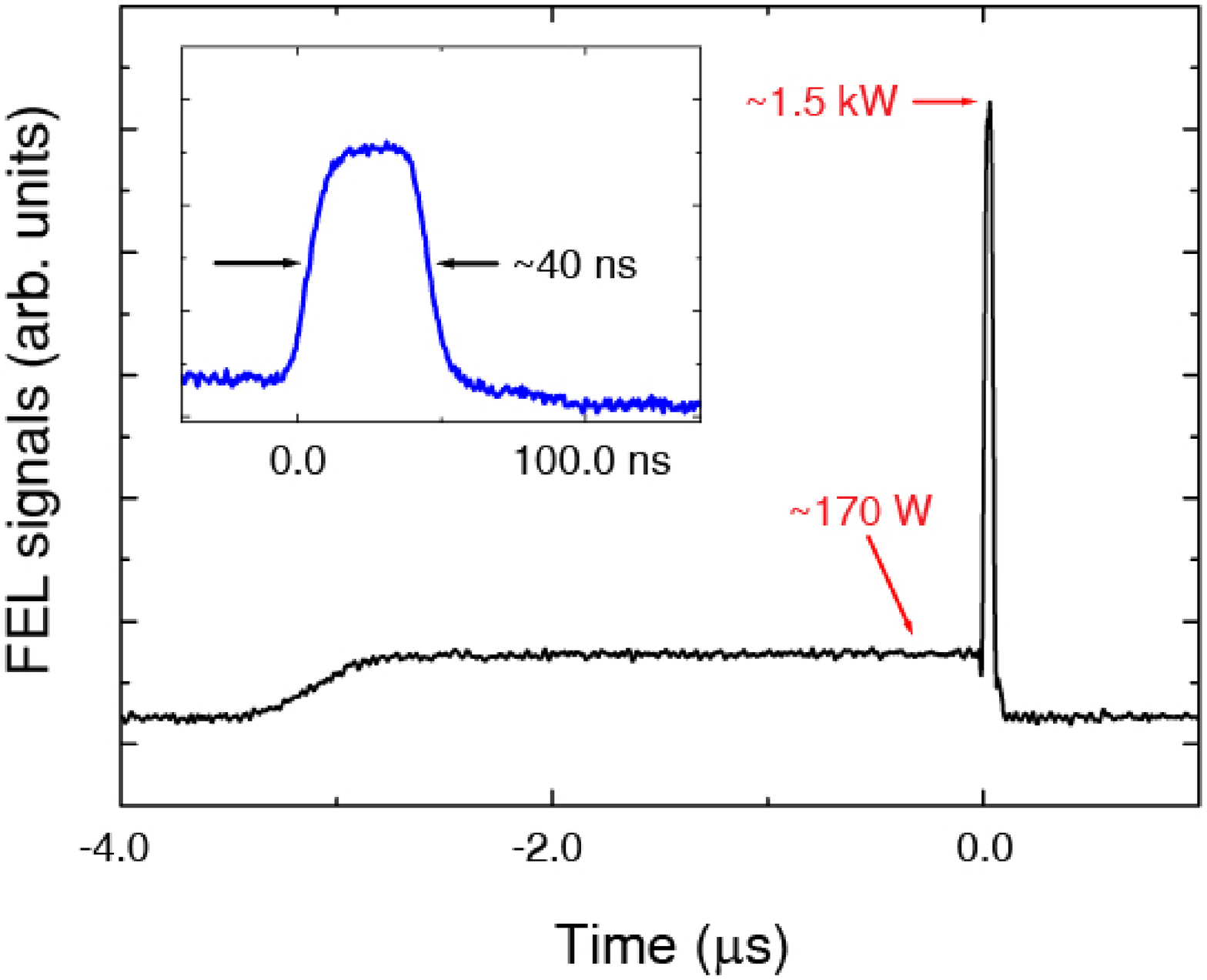}
\caption{\label{figS1} Example of the FEL radiation with the cavity dump coupler (CDC) activated near time 0. The inset shows the FEL CDC output. The length of CDC output is typically $\sim$ 40 ns which corresponds to the round-trip time of the FEL output in the $\sim$ 6 m long FEL cavity. }
\end{figure}

The UCSB free-electron lasers (FELs) are user facilities at the Institute for Terahertz Science and Technology (ITST) at the University of California, Santa Barbara (UCSB). The millimeter wave FEL (mm-FEL) and far infrared FEL (FIR-FEL) are continuously tunable from 0.12-0.89 THz and 0.89-4.7 THz, respectively~\cite{ramian92}. The electron beam, with a several microsecond pulse length, is provided by a 6 MeV, recirculating-beam, electrostatic accelerator. This is a significant difference from most FELs, which use radio-frequency linear accelerators and produce bursts~\cite{felix} or a continuous train~\cite{felbe} of picosecond pulses. The output of the UCSB FEL is a quasi-continuous wave (quasi-cw) beam with $\sim$ kilowatts of peak power and duration comparable to the duration of the electron beam pulse. Therefore, the UCSB FELs are suitable for experiments which require narrow-linewidth radiation at fixed or tunable frequencies~\cite{cole01, carter05, xu07}. For pulsed EPR, at 240 GHz, we have recently implemented injection-locking and cavity-dump coupling (CDC) in the mm-FEL~\cite{takahashi07, takahashi09}. With injection-locking, the linewidth of the FEL is Fourier transform-limited to $<$ 1 MHz. The CDC operation allows us to increase the peak power of the FEL output to over 1 kW with timing shown in Fig.~\ref{figS1}. The FEL output with the CDC lasts $\sim$40 ns, then decays rapidly ({\it i.e.} the FEL output decreases to less than 10 mW within 10 ns).
The FEL output is propagated to the EPR lab with quasi-optical mirrors and polymethylpentene (TPX) lenses in vacuum transport lines. The control system of the FELs has recently been upgraded to a fully computer-controlled system with a friendly graphical user interface (GUI). The program is written in National Instruments Labview.

\newpage
{\bf (ii) Pulse Slicer:}

\begin{figure}
\includegraphics[width=120 mm]{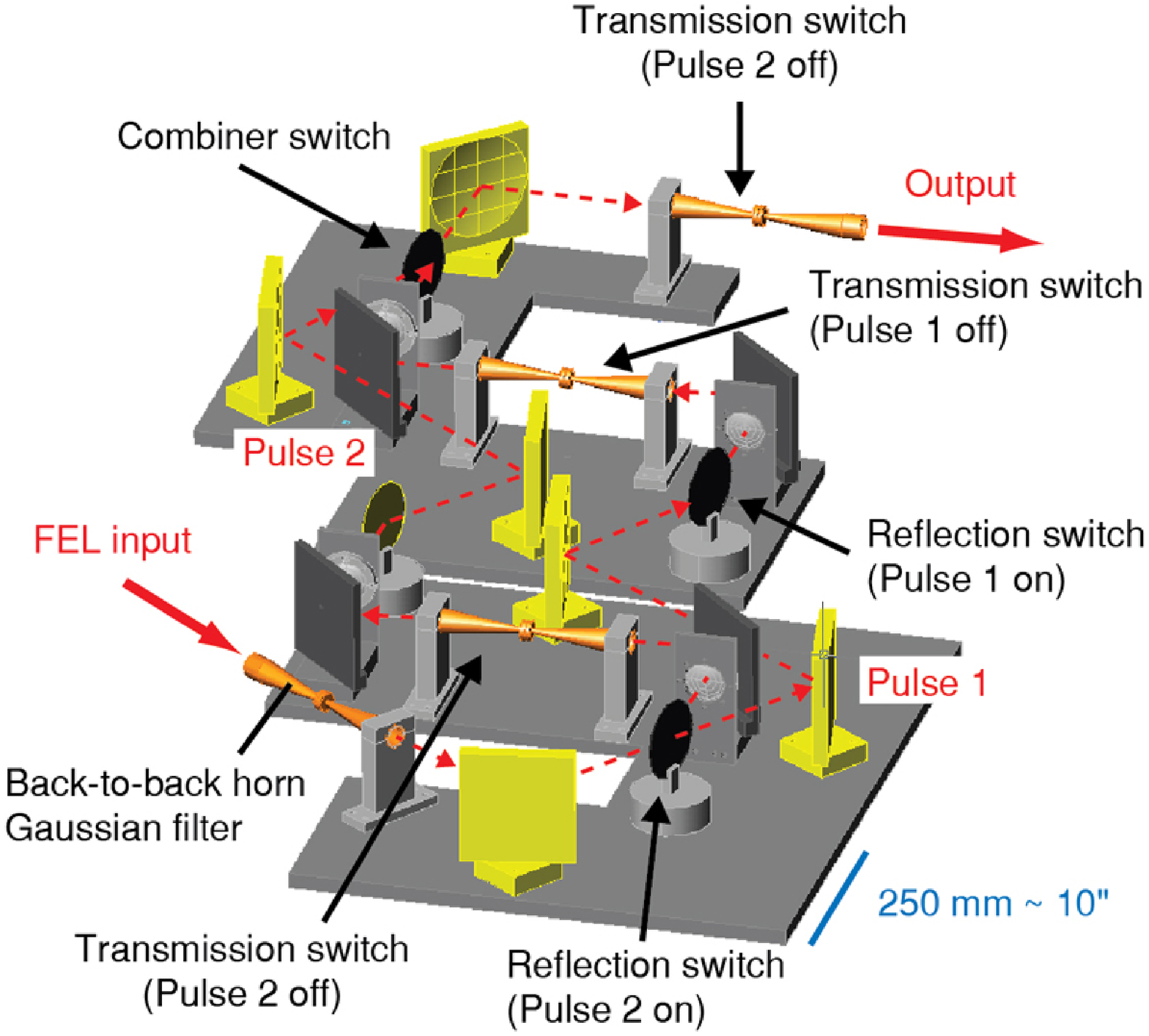}
\caption{\label{figS2} Overview of the pulse slicer system. The pulse slicer system is placed on an optical table. Silicon wafers for the reflection and combiner switches are placed on a rotating stage and are held by a clip. Quasioptical mirrors (gold color) are made out of aluminium and placed on a aluminium plates with dowel pins. There are no angle and position adjustments on the quasioptical mirrors.}
\end{figure}

Fig.~\ref{figS2} shows the overview of the pulse slicer.
The pulse slicer consists of quasioptics, photoactivated silicon switches and Gaussian filters. Quasioptics in the system were designed by a Gaussian mode ray analysis~\cite{kogelik66, goldsmith} and fabricated by the UCSB Physics Department machine shop. We built the quasioptical system as a periodic focusing system with a 500 mm period to cancel any frequency-dependence of the quasioptics~\cite{goldsmith}. As shown in Fig.~\ref{figS2}, we employ a Gaussian filter (back-to-back horn) to spatially filter the FEL at the input of the pulse slicer. In the filter, radiation from the FEL enters a circular corrugated horn which transitions to a rectangular waveguide that supports only a single mode at 240 GHz. A second, identical horn transforms the single mode into a nearly Gaussian beam. The corrugated horn is designed to couple to the fundamental Gaussian mode and to propagate the HE11 mode efficiently (coupling efficiency between the fundamental Gaussian and HE11 modes is $\sim$ 99 $\%$)~\cite{goldsmith}.

\begin{figure}
\includegraphics[width=70 mm]{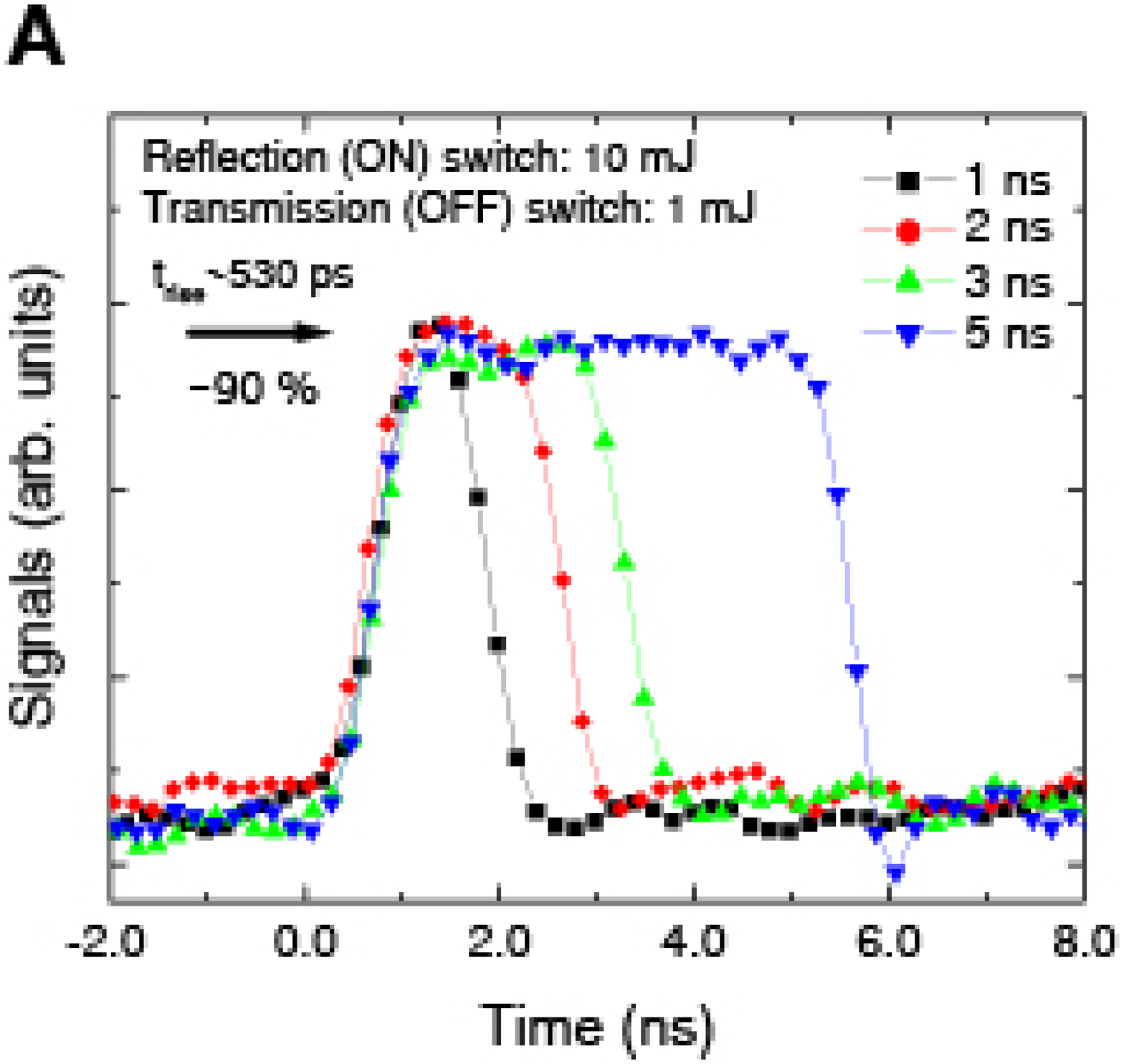}
\includegraphics[width=70 mm]{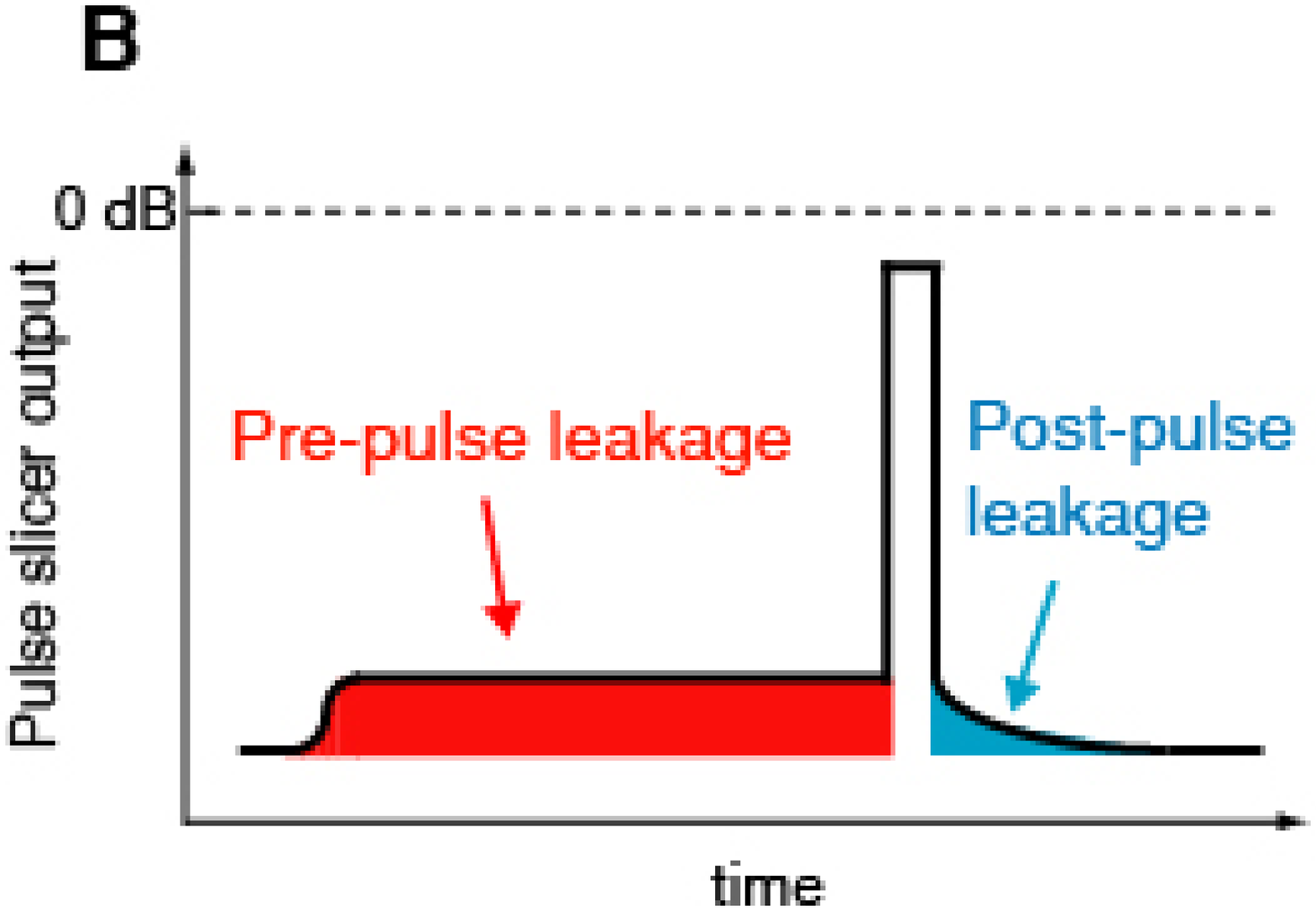}
\includegraphics[width=70 mm]{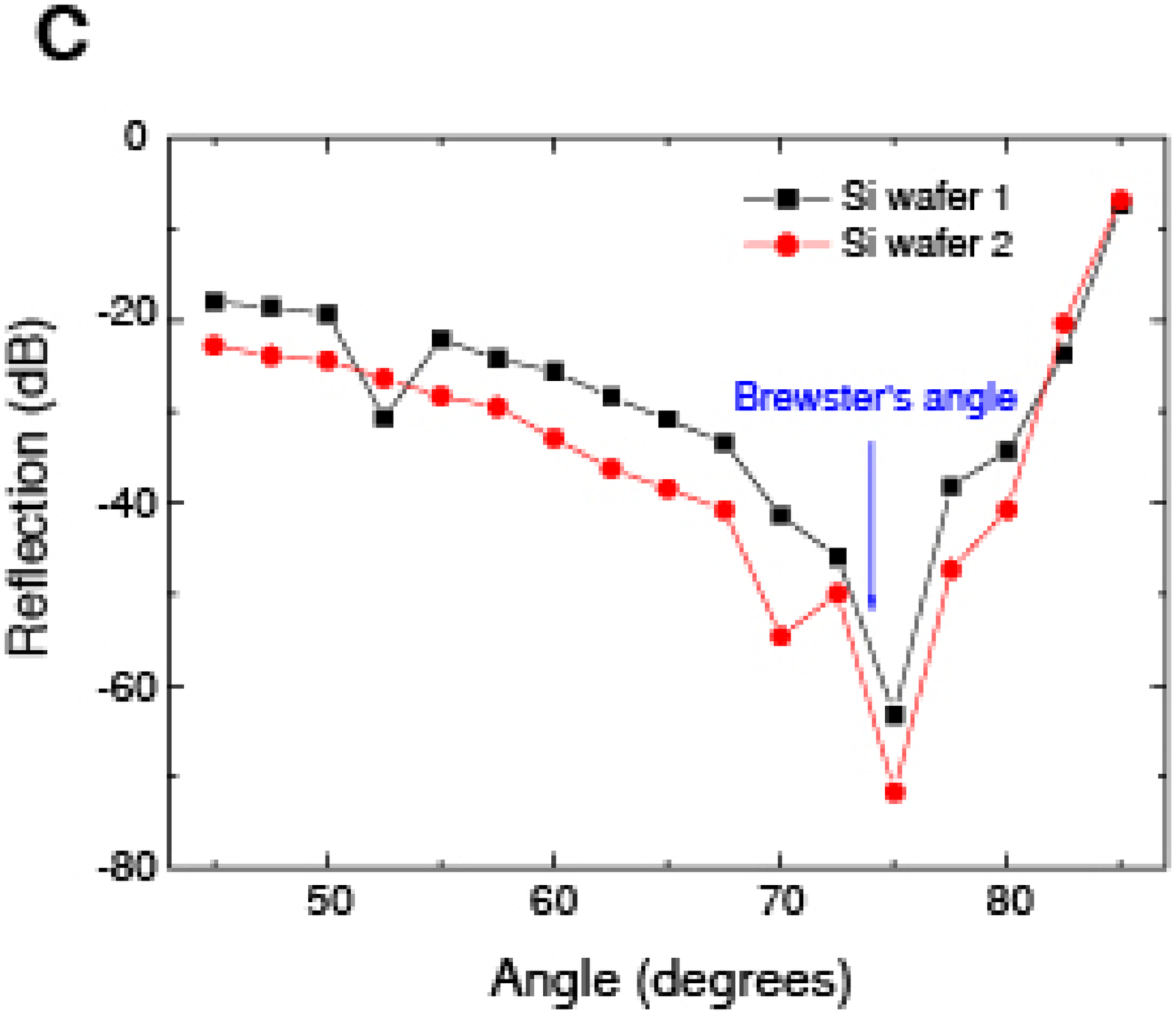}
\includegraphics[width=70 mm]{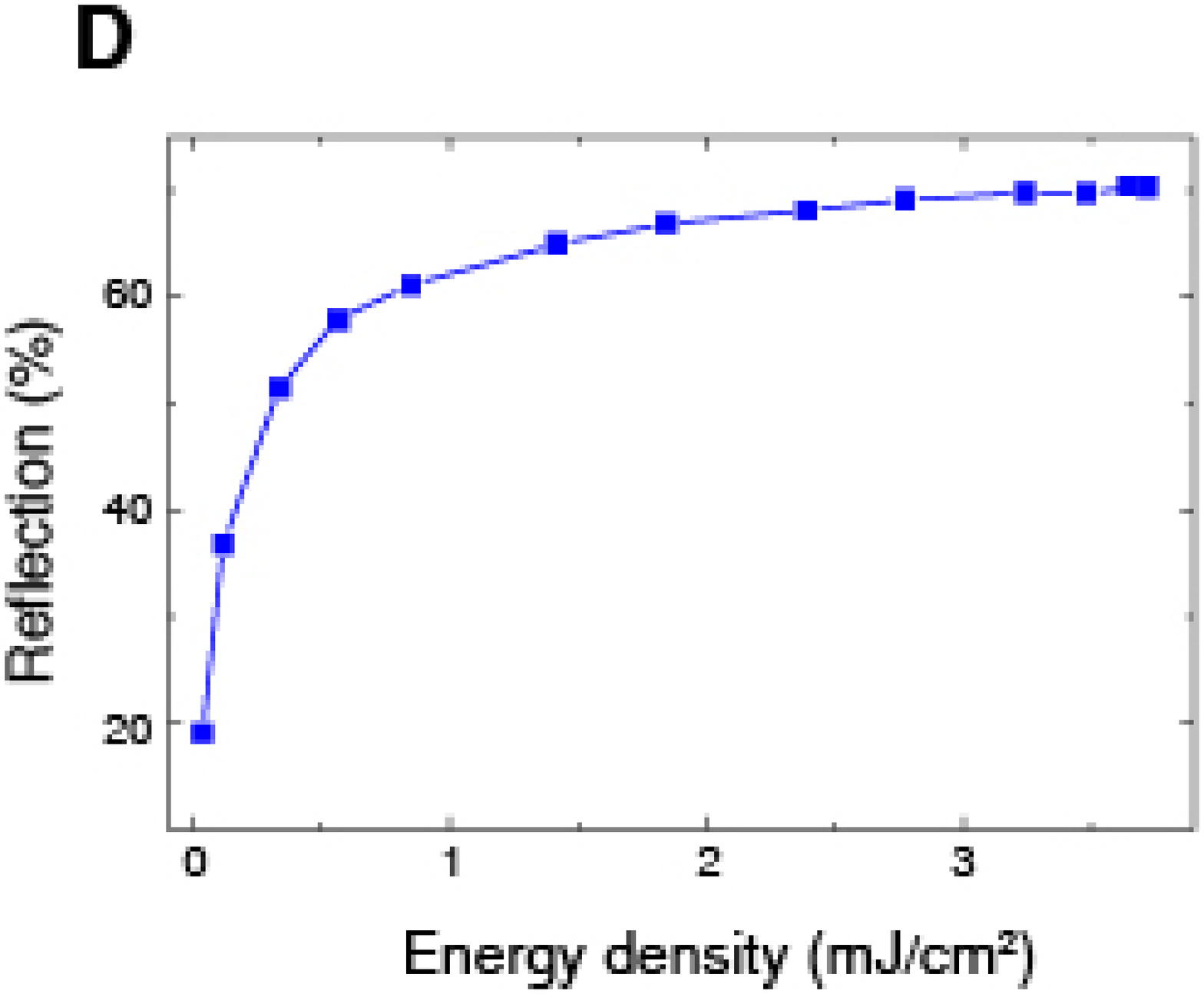}
\caption{\label{FigS3} Performance of the pulse slicer. (a) A single pulse output from the pulse slicer with 1, 2, 3, and 5 ns pulse length. All pulses show a $\sim$ 530 ps rise time. The data shows $\sim$ 90 $\%$ of transmission from the pulse slicer. (b) Schematic of the pulse slicer output to show pre- and post-pulse leakage. Intensity of the leakage is highly exaggerated. (c) The angle dependence of the pre-pulse leakage from reflection switches in their quiescent state. The signals become minimum at Brewster's angle. By using a tuned wafer and nearly Gaussian beams, the minimum reflection is lower than 60 dB. (d) Reflection of a silicon wafer as a function of the energy density of a Nd:YAG laser. Performance of the silicon reflection switches becomes stable (relatively insensitive to pulse-to-pulse fluctuations in Nd:YAG laser fluence) above 1 mJ/cm$^2$.
}
\end{figure}

As shown in Fig.~\ref{figS2}, reflection, transmission and combiner switches in the pulse slicer system are made from photo-activated silicon slabs. The reflectance and transmittance of the Si slabs are controlled by high-energy laser pulses which create a high density of electron-hole pairs on the Si surface~\cite{hegmann96}. A single mm-wave pulse with desirable duration time is sliced by a pair of photo-activated silicon switches, namely reflection and transmission switches, to produce rising and falling edges of the pulse respectively. The rise and fall times of the pulse depend on the switching time of the silicon, which is limited by the performance of a high energy pulse laser. As shown in Fig.~\ref{FigS3}(a), we demonstrated production of pulses down to 1 ns using a high energy laser with a 150 ps pulse length. In the present setup, the pulse slicer consists of two arms to generate two pulses (see Fig.~\ref{figS2}). Each arm has a 100mm ($\sim$4") silicon wafer reflection switch and a transmission switch located inside the back-to-back horn. These switches are activated by high-energy picosecond and nanosecond Nd:YAG lasers. Timing of the activation of the silicon switches is controlled by a digital delay generator (DG 535, Stanford Research Systems) and delay lines built for the Nd:YAG lasers.

Fig.~\ref{FigS3}(b) represents a pulse slicer output. The output signal always contains very small leakage before and after the pulse (pre- and post-pulse leakages). The small leakage is often negligible for conventional low-power pulsed EPR. However, in the case of high-power pulsed EPR, these leakages become problematic, {\it e.g.} by adding background noise and by causing the saturation of EPR. For the present case, the pre-pulse leakage is mainly caused by reflections from reflection switches when the silicon wafers are not activated. In order to minimize the leakage, the slicer plates are mounted at Brewster's angle and made 1/2 wavelength thick. Mounting the plates at Brewster's angle reduces unwanted reflection signals significantly. However, an unacceptable level of leakage remains from a combination of reflections from the front side and back sides~\cite{doty04}, and diffraction of Gaussian waves which makes the incident angle deviate away from Brewster's angle. For further reduction of leakage, we first employ a Gaussian filter and use a large waist size to limit the leakage from diffraction. Second, we optimize the thickness of silicon wafers for the 240 GHz reflection switches to reduce the reflection by destructive interference of the reflection from the front and back sides. Fig.~\ref{FigS3}(c) shows the results of reflection measurements of the tuned silicon wafers taken using a home-built angle dependent measurement system. The reflectance of a single tuned silicon switch is successfully reduced less than 60 dB at Brewster's angle at 240 GHz.

The intensity of the output and the post-pulse leakage depend on the performance of both reflection and transmission switches. As shown in Fig.~\ref{FigS3}(d), we found empirically that a fluence greater than 1 mJ/cm$^2$ at 532 nm wavelength is necessary to saturate the switch. Because of the difference in activation areas for the reflection and transmission switches ($\sim$10 cm$^2$ and \textless 1 cm$^2$ for the reflection and transmission switches respectively), much smaller pulse energies are required for the transmission switch. With optimization, more than 70 $\%$ of the pulse intensity is usually output from the pulse slicer. In combination with the CDC, the post-pulse leakages typically decrease by more than 70 dB within 20 ns after the FEL pulse.

{\bf (iii) EPR Bridge:}

The EPR bridge functions in the FEL-EPR and low-power EPR modes.
The low-power mode, in which a 240 GHz solid-state source with 30 mW output power is placed in the EPR bridge, is used for cw EPR and pulsed EPR. Using the solid-state source, decoherence times $T_2$ for spin-1/2 systems as short as 600 ns are measured. For both the FEL and low-power EPR modes, 240 GHz radiation propagates through quasioptics, and couples to a corrugated circular waveguide which is placed inside a variable temperature cryostat in the room-temperature bore of a superconducting magnet.

\begin{figure}
\includegraphics[width=120 mm]{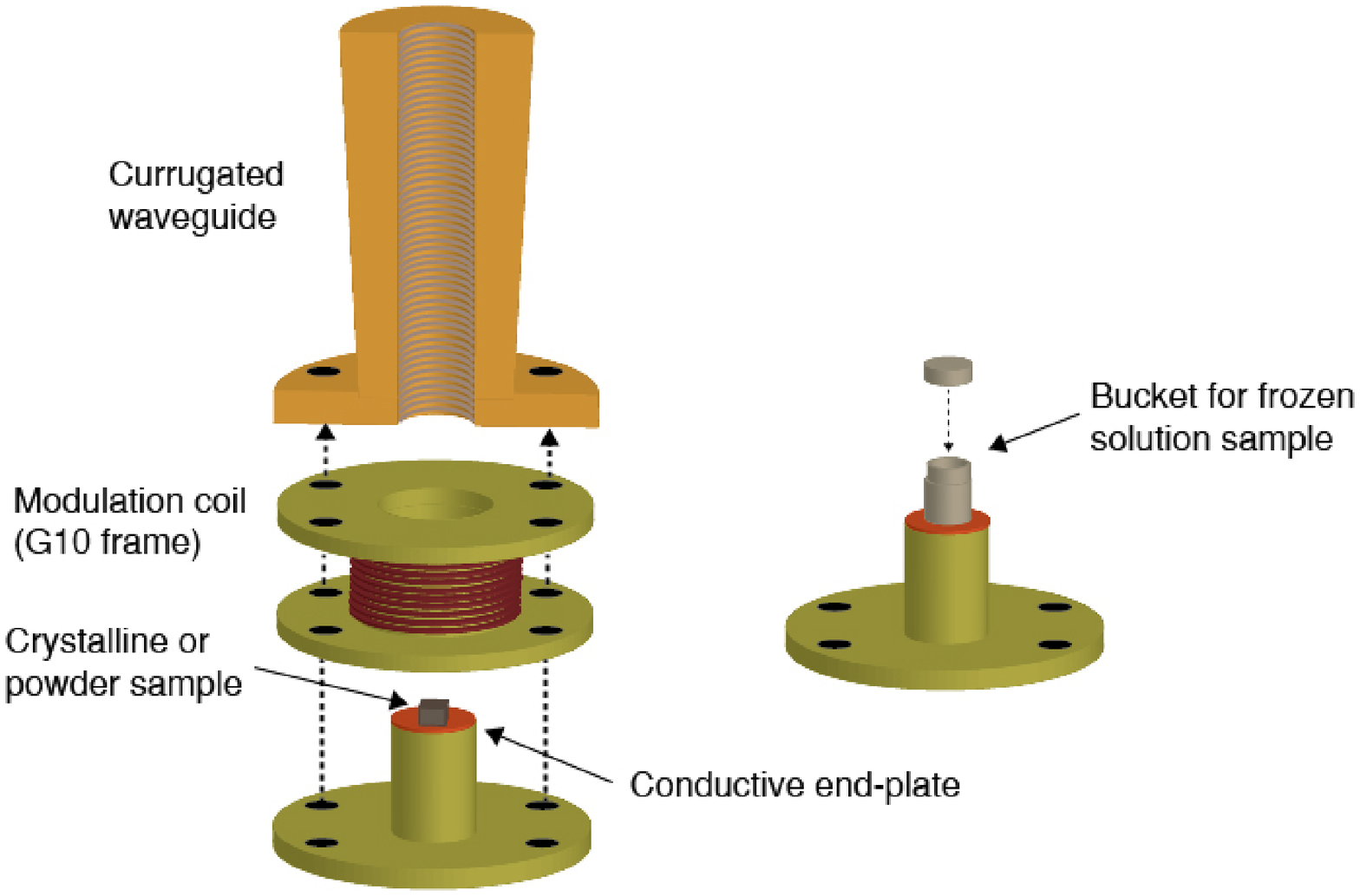}
\caption{\label{FigS4} Schematics of sample holders. Samples are placed at the end of the waveguide. For cw EPR, a modulation coil is placed near the sample holder. Frames of the modulation coil and sample stage are made of G10 fiberglass for low temperature operation.
}
\end{figure}

Samples are placed at the end of the corrugated waveguide. In the current setup, we employ sample holders without a cavity to measure various forms of samples, {\it e.g.} solids, powders and frozen solutions. As shown in Fig.~\ref{FigS4}, a solid or powder sample is directly placed on a conductive end-plate and is located inside the corrugated waveguide or near the end of the waveguide. For frozen solution samples, we use a "bucket" with a volume of 20 microliters to hold the solution sample. The bucket is made of teflon which is placed on the end-plate.

\begin{figure}
\includegraphics[width=150 mm]{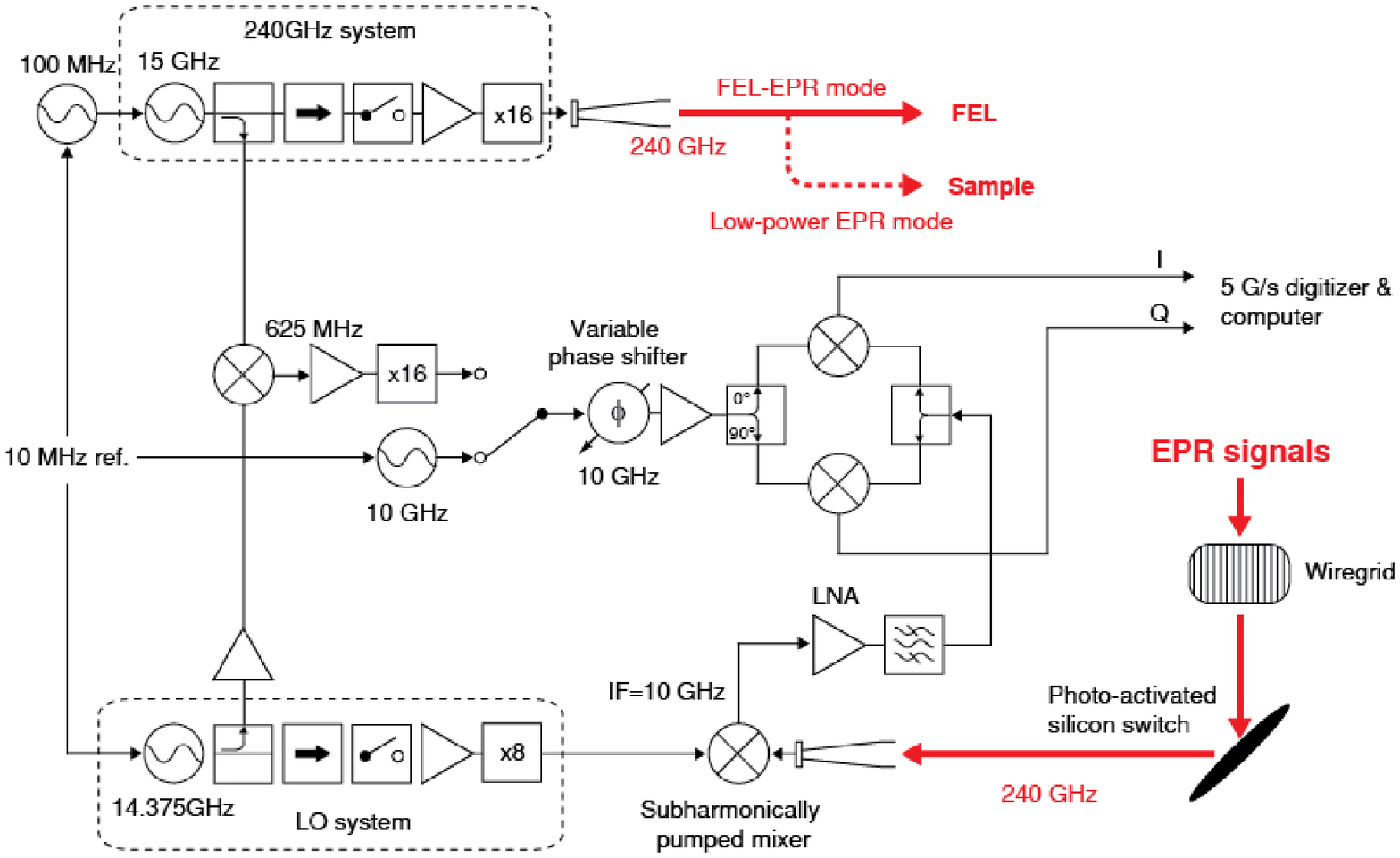}
\caption{\label{FigS5} Schematic of the detection system. The solid-state 240 GHz system (Virginia Diode Inc.) consisting of a 15 GHz synthesizer, directional coupler, isolator, p-i-n switch, amplifier and frequency multipliers is used as an injection source in the FEL-EPR mode and as a source for cw and pulsed EPR spectroscopy in the low-power EPR mode. The reference at 10 GHz is produced by mixing 15 GHz from the source and 14.375 GHz from the LO, then by multiplying 625 MHz by 16 times using frequency multipliers (Miteq). The 10 GHz reference is split into 0 and 90 degree reference using 90 degree hybrid 3 dB directional coupler (Advanced Technical Materials Inc.), and then is mixed with the signal to produce 0-500 MHz in-plane and quadrature signals.
}
\end{figure}

Fig.~\ref{FigS5} shows our detection system. In order to isolate EPR signals from FEL leakage, and to protect the mixer detector from the high power FEL pulses, we employ a two-stage isolation system. The first stage is a wiregrid polarizer (Thomas-Keating Ltd.). The wiregrid polarizer reflects 240 GHz waves when their E-field components are parallel to the direction of the wires. Because the reflected pulses have very small cross-polarized components while half of EPR circularly polarized signal is cross-polarized, the wiregrid can separate EPR signals well from exciting pulses. This isolation is typically more than 25 dB. The second stage is a photo-activated silicon switch. As shown in Fig.~\ref{FigS3}(c) and (d), this isolation is more than 60 dB. The combination of these isolation systems keeps the background levels sufficiently low for the experiments presented in the paper.

Our detection system is based on a superheterodyne receiver. Using a subharmonically pumped mixer and the local oscillator (Virginia Diode, Inc.), signals at 240 GHz are converted into signals at 10 GHz, then signals are amplified by a low noise amplifier (noise figure=0.9 dB, Miteq). The reference at 10 GHz is produced either by a synthesizer (Microlambda wireless) or by a home-built reference circuit including a x16 frequency multiplier for the FEL-EPR and low-power EPR modes respectively. Then, the signal and reference at 10 GHz are mixed down into a band stretching from 0-500 MHz. The noise temperature of the detection system is $\sim$1300 K. Both in-plane and quadrature signals are detected by a digitizer with 5 Gs/s sampling rate and 1 GHz bandwidth and are recorded by a computer.

Our pulsed EPR setup employs a 12.5 tesla superconducting magnet (Oxford Instruments). The magnet consists of 12.5 T main coil and $\pm$0.06 T sweep coil and 89 mm room temperature bore. Its stray magnetic field is actively shielded to not have any effects on the UCSB FEL. For low temperature measurements, we employ a flow cryostat (Janis Research Co.) with 62 mm inner diameter. The cryostat has a quartz window at the bottom of the cryostat for access to optical excitations.
